\documentclass[11pt]{article}

\usepackage{epsfig}
\usepackage{amsmath}
\usepackage{soul,color}
\usepackage{bm}
\usepackage{epsfig}
\usepackage{natbib}
\usepackage{hyperref}
\usepackage{multirow}
\newcommand{\tr}{^{\prime}}
\def\b#1{\mbox{\boldmath $#1$}}    

\newcommand{\al}{\alpha}
\newcommand{\be}{\beta}
\newcommand{\de}{\delta}
\newcommand{\la}{\lambda}

\newcommand{\ga}{\gamma}

\usepackage{bm}
\usepackage{epsfig}

\begin{document}
\title{\vspace*{-1cm}A multidimensional latent class IRT model\\
for non-ignorable missing responses\footnote{The present paper has been accepted for the publication on {\em Structural Equation Modeling: A Multidisciplinary Journal.}}} 
\author{Silvia Bacci\footnote{Both authors acknowledge the financial support from the grant FIRB (``Futuro in ricerca'') 2012 on ``Mixture and latent variable models for causal inference and analysis of socio-economic data'', which is funded by the Italian Government (RBFR12SHVV).
The authors are also grateful to Dr. B. Bertaccini of the University of Florence (IT) for making available the data.}\\
{\em Department of Economics, University of Perugia (IT)}\\
{\em email}: silvia.bacci@stat.unipg.it\vspace*{5mm} \and
Francesco Bartolucci$^*$\\
{\em Department of Economics, University of Perugia (IT)}\\
{\em email}: bart@stat.unipg.it\vspace*{5mm}
}   \maketitle

\def\baselinestretch{1}
\begin{abstract}\noindent
We propose a structural equation model, which reduces to a multidimensional latent 
class item response theory model, for the analysis of 
binary item responses with non-ignorable missingness. 
The missingness mechanism is driven by two sets of latent 
variables: one describing the propensity to respond 
and  the other referred to the abilities measured by the test items.
These latent variables are assumed to have a discrete distribution,
so as to 
reduce the number of parametric assumptions regarding the latent structure of the model.
Individual covariates may also be included through a multinomial logistic parametrization of the probabilities of each support point of the distribution of the latent variables. 
Given the
discrete nature of this distribution, the proposed model is
efficiently estimated by the Expectation-Maximization algorithm.
A simulation study is performed 
to evaluate the finite-sample properties of the parameter estimates. Moreover, 
an application is illustrated to data coming from a
Students' Entry Test for the admission to some university courses.


\end{abstract}
\newpage

\section{Introduction} \label{sec:1}\vspace*{-0.25cm}
A relevant problem in applications of Item Response Theory (IRT) models
is related to missing responses to some items.
Following the general theory of \cite{little:rubin:2002}, we define missing item responses to be {\em ignorable} if: ({\em i}) these responses
are missing at random (MAR), that is, the event that the response to an item is missing is conditionally independent of the (unobservable) response to this item given the observed responses to the other items and ({\em ii}) the missing mechanism is governed by a model based on a distinct set of parameters with respect to the parameters of the model governing the response process. Under ignorability, 
maximum likelihood estimation of the parameters of the IRT model of interest is only based on the observed responses.

Obviously, when condition ({\em i}) or ({\em ii}) above does not hold, then missing
responses are \textit{non-ignorable} and the missingness mechanism must be modeled along with the relationships of direct interest to avoid wrong inferential conclusions
and loss of relevant information for the assessment of the examinees' ability level. 
A typical example of non-ignorable missing responses (or missing not at random, 
MNAR) is observed with educational  
tests where, in order to avoid guessing, a wrong item response is penalized
to a greater extent in comparison with 
a missing response. In such a context, it is natural
to suppose that the choice of not answering to a given item is related to the 
ability (or abilities) measured by the test.  

In the statistical literature there exist different approaches to 
model a non-ignorable missing mechanism. Among the best known, we recall the {\em selection approach} \citep{diggle:ken:94} and the {\em pattern-mixture approach} \citep{little:93}. 
The first formulation defines a joint model of observed and missing 
responses and factorizes 
the corresponding distribution in  
a marginal distribution for the complete data (union of 
observed and missing responses) and a conditional distribution for the missing data given the 
complete data. In contrast, the pattern-mixture approach specifies the marginal 
distribution for the missing responses and the conditional distribution of the complete data
given the missing responses. 

Recently, \cite{for:07b} showed that on the basis of the latent class (LC) model 
\citep{laza:henr:68,good:74} it is possible to define class of MNAR models, which is distinct from that of selection models and that of pattern-mixture models.
He treated the presence of non-ignorable missing entries in the  case of repeated measurements of the same variable or observations of different variables made on the same individuals. 
The approach is based on creating an extra category for missing responses, so as to model the missingness mechanism, and analyzing the data coded in this way by means of an LC model.
In this way each latent class is also characterized in terms of missing responses.
Moreover, individual covariates may influence the class weights, so that the latent class distribution becomes individual-specific as in the latent regression models of \cite{roche:miglio:97} and \cite{bart:forc:06}. 

In the statistical literature, \cite{harel:schafer:09} proposed the use of LC models to treat cases where missingness is only partially ignorable. 
They introduced the concepts of {\em partial ignorability}, which supposes that a summary of 
the missing-data indicators depends on the missing values, and of {\em latent ignorability}, 
based on assuming that the missing-data indicators depend on a summary of the missing values. 
In the context of item responses, they proposed to create a binary 
missingness indicator corresponding to each item and to fit an LC model treating 
these indicators as additional items.
In this way, each latent class does not only summarize answers to questionnaire items, but  
the individual propensity to answer.

In the IRT context,  
which is  
of our specific interest, several approaches may be adopted to deal with MNAR responses. 
The most naive ones consist of adopting simple IRT models that ignore the missing responses or consider the omissions as wrong responses. 
\cite{brad:tho:98} and \cite{rose:10} warned against the drawbacks of these approaches, 
which lead to biased estimates of model parameters and, therefore, to unfair comparisons 
between persons. 
To overcome these limitations, several authors proposed approaches based on modeling the non-ignorable missingness process. 
\cite{mou:knott:00} and \cite{mou:muir:00}, among others, 
discussed a nominal IRT model, with possible covariates for the 
ability, where the missing responses are treated as separate response categories, elaborating an original idea by \cite{bock:72}.
On the other hand, \cite{rose:10} proposed a latent regression model 
 where the latent ability is regressed on the observed response rate, referring in this way the missingness mechanism to the covariates rather than to the responses. 

An interesting stream of research has been introduced by \cite{lord:83}, who suggested to treat the problem of 
MNAR responses by assuming that the observed item 
responses depend both on the latent ability (or abilities), intended to be measured by the test, and on another latent variable which represents the ``temperament'' of respondents, and describes their propensity to respond. 
Elaborating the approach of \cite{lord:83}, \cite{hol:glas:05} 
discussed a unified model-based approach for handling non-ignorable missing data and, therefore, assessing the extent to which the missingness mechanism may be ignored. The adopted approach relies on multidimensional IRT models \citep{reckase:10} and 
on the assumption that the latent traits are normally distributed. 

It is also worth recalling the work of \cite{bertoli:punzo:12} that proposed an alternative non-parametric approach based on the conditional maximum likelihood estimation method, 
where a  multidimensional IRT model is specified according to the Rasch model 
assumptions \citep{rasch:60}.
The main drawback of the conditional approach is that it does not allow us to measure the correlation between the assumed 
latent variables; moreover, its use is limited to settings for which the Rasch model is realistic. 

The above mentioned approaches based on the introduction of a latent variable describing the tendency to respond are well suited to a Structural Equation Model (SEM) formulation \citep{gold:72, duncan:75, bol:rabe:skr:08}, which allows for several types of generalizations (e.g., semiparametric specification of the latent trait distribution and effect of individual covariates on the latent traits).

Aim of the present article is to introduce a SEM, which reduces to a special type of  multidimensional LC IRT model, to deal with non-ignorable missing responses to a set of 
test items. 
The model is based on the assumption of discreteness of the latent variables, not only for the response process but also for the missingness process. Therefore, with respect to traditional SEM, the proposed model takes the form of a finite mixture SEM 
 \citep{jed:jag:des:1997, dol:maas:98, arm:stein:witt:99}.

The basic model we rely on was introduced by \cite{bart:07} and it is based on two main assumptions:
({\em i}) more latent traits can be simultaneously considered and each item is associated with only one of them  \citep[between-item multidimensionality, see][]{ada:wil:wan:97},
 and 
({\em ii}) these latent traits are represented by a
random vector with a discrete distribution common to all subjects, so that each support point of such a distribution identifies a different latent class of individuals having homogenous unobservable characteristics.
Moreover, with binary response variables, either a Rasch or a two-parameter logistic (2PL) parametrization \citep{birn:68} may be adopted for the probability of a correct response to each item.
In this context,
we propose to include a further discrete latent variable to model the probability of observing a response to each item, so that the non-ignorable missing process may be treated in a semiparametric way,  as made for the response process.
Other than extending the model of  \cite{bart:07} to allow for missingness, we also extend it to allow for latent individual covariates which may explain the probability of belonging to a given latent class. 

The approach proposed in this article joins the latent class approach of \cite{for:07b} with the parametric approach of \cite{hol:glas:05} developed in the IRT setting.
Several advantages with respect to the last one may be found. 
First, the proposed model is more flexible because it does not  introduce any parametric assumption about the distribution of the latent variables. 
Second, detecting homogenous classes of individuals is convenient for certain decisional processes, because individuals in the same class may be associated to the same decision (e.g., students admitted, admitted with reserve, not admitted to university courses). 
Finally, our model allows us to skip the well-known problem of the intractability of multidimensional integrals which characterizes the marginal log-likelihood function of a continuous  multidimensional IRT model. 
Indeed,  
parameter estimation may be performed through the discrete marginal maximum likelihood method, based on an Expectation-Maximization (EM) algorithm \citep{demp:laird:rub:77}, and implemented in an {\tt R} function that we make publicly available.  

In order to assess the finite-sample properties of the parameter estimates obtained from the EM algorithm, we have performed a simulation study under different scenarios corresponding to different structures of missing data.
In this way, we can also assess the impact of missing data on the quality of the parameter estimates with respect to the case in which all data are observed. The proposed approach is also illustrated through an application to real data coming 
from the Students' Entry Test given at the Faculty of Economics of an Italian university in 2011. 
The test is composed of 36 multiple-choice items devoted to measure three latent abilities (Logic, Mathematics, and Verbal comprehension) and certain covariates 
are also included. 

The remainder of the paper is organized as follows. 
We first describe the proposed structural model to account for the presence of non-ignorable missing responses in the IRT context and its statistical formulation. 
Then, some details about the estimation procedure through the EM algorithm are described together with other details about likelihood inference. 
In the sequel, we illustrate the simulation study to evaluate the adequacy of the proposed approach.
The application of the proposed approach to the data arising from the Students' Entry Test is illustrated in the last section. 
\section{Proposed SEM formulation}\label{sec:mod} 
In this section, we describe the proposed approach to model MNAR item responses \citep{little:rubin:2002}.
We begin by illustrating the proposed SEM and then we provide the resulting statistical formulation which may be cast in the class of multidimensional IRT models.
\subsection{Structural model}\label{sec:sem}\vspace*{-0.25cm}
For a random subject drawn from the population of interest, denote by $Y_j$ the response provided by the subject to
binary item $j$, with $j=1,\ldots,m$. In order to model the response process, we have to consider that the subject may answer correctly ($Y_j = 1$) or incorrectly ($Y_j=0$) or he/she may skip the question, so that 
$Y_{j}$ can be observed or not. Therefore, 
for $j=1,\ldots,m$, we also introduce the binary indicator $R_{j}$ equal to 1 if the  
individual provides a response to item $j$ and to 0 otherwise (i.e., $Y_{j}$ is missing); 
see also 
\cite{harel:schafer:09}. Moreover, we consider a set of $c$ exogenous individual covariates denoted by $X_1,\ldots,X_c$.

In order to explain the association between the 
exogenous variables $X_1,\ldots,X_c$ and the endogenous variables $Y_1,\ldots,Y_m$, we introduce two latent variables.
The first of these latent variables, denoted by $U$, represents the latent trait that is measured by the test items (e.g., ability in Mathematics). The second latent variable, denoted by $V$, is interpreted as the propensity to answer \citep[as in][]{lord:83}, the opposite of an {\em aversion to risk} if a wrong response is somehow penalized. 
Based on these latent variables and considering 
$Y_j$ and $R_j$ 
as deriving from a discretization of continuous variables denoted by $Y^*_j$ and $R_j^*$, 
we formulate the following  
equations entering the \textit{measurement component} of the proposed SEM:
\begin{eqnarray}
Y_j&=&I\{Y^*_j\geq 0\},\label{eq:11}\\
R_j&=&I\{R^*_j\geq 0\},\label{eq:12}\\
Y_j^*&=&\al_jU-\be_j+\varepsilon_{1j},\label{eq:13}\\ 
R_j^*&=&\ga_{1j}U+\ga_{2j}V-\de_j+\varepsilon_{2j},\label{eq:14} 
\end{eqnarray}
for 
$j=1,\ldots,m$, where $I\{\cdot\}$ is the indicator function equal to 1 if its argument is true and to 0 otherwise and $\varepsilon_{1j}$ and $\varepsilon_{2j}$ are independent error terms. Moreover, 
the slope  $\al_j$  measures the effect of an increase of the 
latent variable $U$ on $Y_j^*$ and, similarly, 
$\ga_{1j}$ and $\ga_{2j}$ measure the effect on $R_j^*$ of $U$ and $V$, respectively.

According to the proposed model, the observed response to a given item $j$ depends only on the latent ability $U$ measured by the test, whereas the event of answering to item $j$ depends both on 
$U$ and on the propensity to respond $V$. 
Therefore, provided that $\ga_{2j}>0$, $R_j^*$ tends to increase  
with the propensity to respond  given the latent ability level.
Similarly, provided that $\ga_{1j}>0$,  
$R_j^*$ tends to increase 
with the ability level even if the propensity to answer remain constant. 
The idea behind this assumption is that better students are more willing to respond due to their confidence on 
the correctness of the response.
Note that the adopted formulation reminds model G3 of \cite{hol:glas:05}, whereas model G2 proposed by the same authors is obtained by imposing the constraint $\ga_{1j}=0$, $j=1,\ldots,m$, which implies the absence of any direct effect of $U$ on $R_j^*$ and, therefore, denotes that the missingness process may be ignored. 
Finally, $\be_j$ and $\de_j$ denote other two parameters characterizing  item $j$, which may be interpreted as {\em difficulty parameters} because higher values of them correspond to smaller values of $Y_j^*$ and $R_j^*$. 

The proposed SEM formulation is completed by assuming that: (\emph{i}) the latent variables $U$ and $V$ are conditionally independent given the covariates $X_1,\ldots,X_c$  and that (\emph{ii}) a direct effect of these covariates on the response variables is ruled out. How we formulate the conditional distributions of $U$ and $V$ given the covariates 
will be clarified in the following section.

\begin{figure}[!ht]
\begin{center}
\caption{\em Proposed SEM for only one covariate ($c=1$), four items ($m=4$), and two dimensions ($s=2$), with the first two items measuring the first dimension and the other two items measuring the second dimension. \vspace{5mm}
}\label{fig2}
\includegraphics{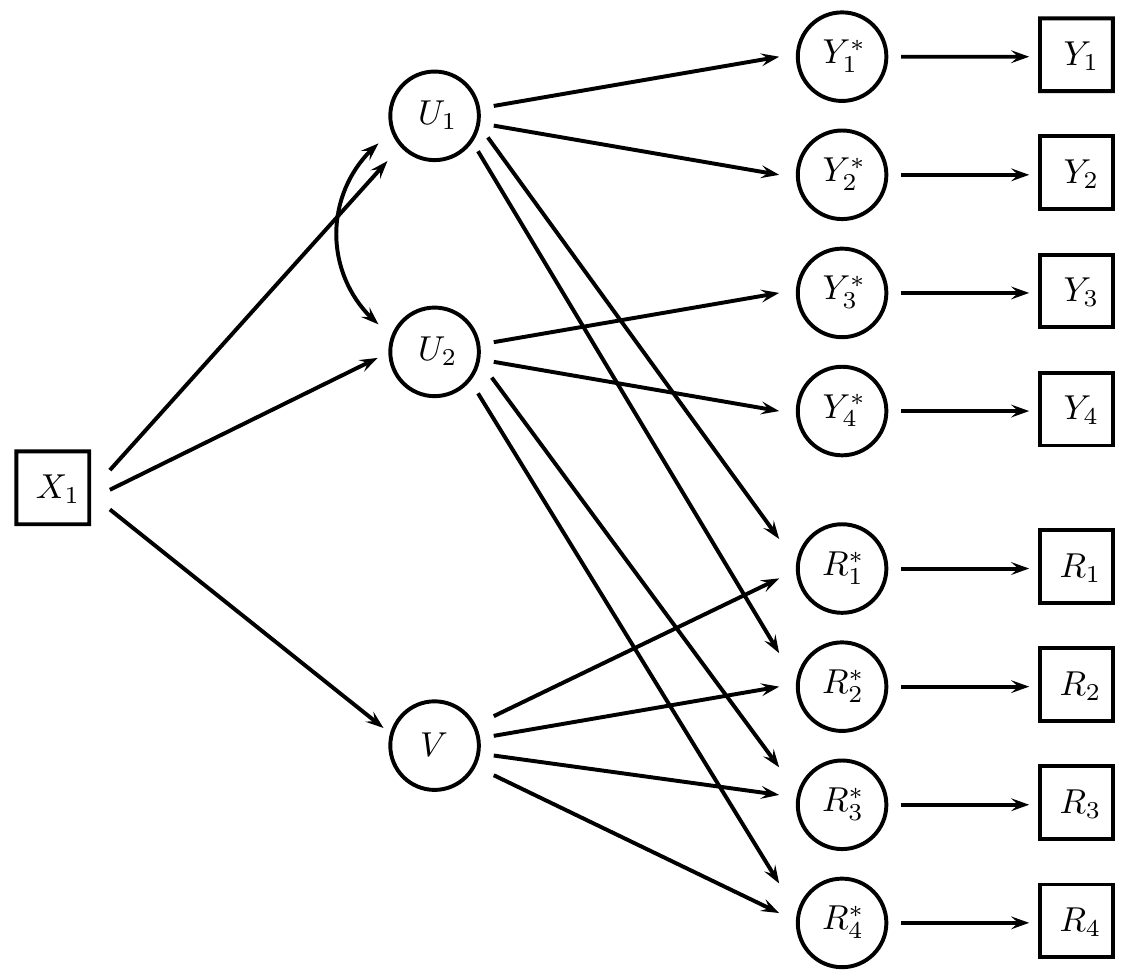}
\end{center}
\end{figure}

The above approach  may be easily extended to the multidimensional case 
with items measuring $s$ different latent traits 
(e.g., ability in Mathematics, ability in Logic, ability in Verbal comprehension),
which are represented by the latent variables $U_1,\ldots,U_s$,
assuming, in addition to (\ref{eq:11}) and (\ref{eq:12}), that
\begin{eqnarray}
Y_j^*&=&\al_j\sum_{d=1}^s z_{dj}U_d-\be_j+\varepsilon_{1j},\label{eq:23}\\ 
R_j^*&=&\ga_{1j}\sum_{d=1}^s z_{dj}U_d+\ga_{2j}V-\de_j+\varepsilon_{2j}, \label{eq:3}
\end{eqnarray}
for $j=1,\ldots,m$.
In comparison with the structural model based on equations (\ref{eq:11})-(\ref{eq:14}), the new one changes in the last two equations 
involving the indicator variables $z_{dj}$, which are equal to 1 if item $j$ measures latent trait of type $d$ and to 0 otherwise. 
A between-item multidimensional approach 
 \citep{ada:wil:wan:97}  is assumed with reference to the measurement of the $s$ latent abilities, indicating that each item measures only one of them. 
On the other hand, a within-item multidimensional approach is here adopted for the indicator $R_j$, since it is affected by two latent variables.
In any case, our conceptual model still assumes 
one latent variable $V$ for the propensity to answer (for an illustration see Figure \ref{fig2}). A possible alternative, which is more complex to deal with, is represented by the introduction of $s$ latent variables $V_1, \ldots, V_s$ for the propensity to answer, one for each latent variable $U_1, \ldots, U_s$.\vspace{5mm}

FIGURE \ref{fig2} HERE \vspace{5mm}
\subsection{Statistical model}\label{sec:statmod}
In order to estimate the structural model outlined above, we need to formulate
assumptions on the conditional distribution of the latent variables given the
covariates and on the distribution of the error terms $\varepsilon_{1j}$ 
and $\varepsilon_{2j}$ in the equations for $Y_j^*$ and $R_j^*$.

Regarding the first aspect, we assume that the vector of latent variables $(U_1, \ldots, U_s)\tr$ and $V$ are independent, given the covariates  
$X_1,\ldots,X_c$, and have a discrete distribution, with $k_1$ support points $(u_{1h_1}, \ldots, u_{sh_1})\tr$, $h_1=1,\ldots,k_1$, and $k_2$ support points $v_{h_2}$, $h_2=1,\ldots,k_2$, respectively. The corresponding weights are denoted by
\begin{eqnarray*}
\la_{h_1}(\b x) &=& p(U_1 = u_{1h_1},\ldots,U_s=u_{sh_1}|X_1 = x_1,\ldots,X_c=x_c),\quad h_1=1,\ldots,k_1,\\
\pi_{h_2}(\b x) &=& p(V=v_{h_2}|X_1 = x_1,\ldots,X_c=x_c), 
\quad h_2=1,\ldots,k_2,
\end{eqnarray*}
which depend on the column vector of observed individual covariates 
$\b x=(x_1,\ldots,x_c)\tr$. 
This is equivalent to assuming that individuals come from 
latent classes which are internally homogenous in terms of the latent traits
measured by the questionnaire.
Note that, with respect to traditional LC models, there is not a set of weights 
common to all subjects in the sample, but weights $\la_{h_1}(\b x)$ and $\pi_{h_2}(\b x)$ which are subject-specific, as they depend on the individual covariates.
In particular, we assume the following multinomial logistic parametrization, which is similar to that adopted by \cite{for:07b}:
\begin{equation}
\log{\frac{\la_{h_1+1}(\b x)}{\la_1(\b x)}} = 
\phi_{0h_1} + \b x\tr\b\phi_{1h_1},  \quad h_1 = 1, \ldots, k_1-1,
\label{eq:cov}
\end{equation} 
where $\phi_{0h_1}$ and $\b\phi_{1h_1}$ are regression parameters to be estimated. A similar parametrization,
based on regression parameters $\psi_{0h_2}$ and $\b\psi_{1h_2}$, is assumed for the  
conditional probabilities $\pi_{h_2}(\b x)$, $h_2=1,\ldots,k_2-1$.

Regarding the error terms $\varepsilon_{1j}$ and $\varepsilon_{2j}$, we assume that
they are independent and have a standard logistic distribution.
This is a convenient assumption implying that the distribution of $Y_j$ and $R_j$ given the corresponding latent variables satisfies a logistic model. 
In particular, with 
\begin{eqnarray*}
p_{h_1j}&=&p(Y_{j} = 1|U_1=u_{1h_1},\ldots,U_s=u_{sh_1}),\quad h_1=1,\ldots,k_1\\
q_{h_1h_2j}&=&p(R_{j} = 1|U_1=u_{1h_1},\ldots,U_s=u_{sh_1},V=v_{h_2}),\quad h_2=1,\ldots,k_2,
\end{eqnarray*} 
we have the following 2PL
parametrization \citep{birn:68}:
\begin{eqnarray}
\log\frac{p_{h_1j}}{1-p_{h_1j}} &=& 
\al_j \sum_{d=1}^s z_{dj}u_{dh_1} - \beta_j, 
\quad j=1,\ldots,m,\label{eq:par1}\\ 
\log\frac{q_{h_1h_2j}}{1-q_{h_1h_2j}}&=&  
\ga_{1j}\sum_{d=1}^s z_{dj}u_{dh_1} + \ga_{2j}v_{h_2} - \de_j, \quad j=1,\ldots,m.
\label{eq:par2}
\end{eqnarray} 
Using the terminology of IRT models,  
$\al_j$, $\ga_{1j}$, and $\ga_{2j}$ are {\em discriminating parameters} and 
$\beta_j$ and $\de_j$ are {\em difficulty parameters}. In particular, $\al_j$
measures the discriminating power of item $j$, that is, how the probability
of responding correctly to the item depends on the latent ability, whereas
$\ga_{1j}$ and $\ga_{2j}$ measure how the probability of responding (correctly or not) to item $j$ depends on both latent ability and  propensity to answer, respectively. 
A more parsimonious model is obtained by constraining all the discriminating parameters 
to be equal one other, that is, $\al_j=1$, $\ga_{1j}=1$, and $\ga_{2j}=1$ for  
$j=1, \ldots, m$. In this way a Rasch type model \citep{rasch:60} is specified.

The parametrization illustrated above may be extended to take into account general situations, such as that of differential effects of $U_d$ on $R_j^*$ across the latent classes $h_1=1, \ldots, k_1$.
For this aim, we should introduce a discriminating parameter depending on the specific latent class; therefore we have to substitute $\ga_{1j}$ with $\ga_{1h_1j}$ in equation (\ref{eq:par2}). 
However, generalizations of this type give rise to more complex models, the interpretation and estimation of which may be complicate.

The assumptions formulated in the previous section imply an assumption that
in the IRT literature is known as {\em local independence}; see, for instance,
\cite{ham:swam:85}. In the present context, 
this assumption means that all variables $Y_1,\ldots,Y_m$
and $R_1,\ldots,R_m$ are conditionally independent given the latent vector 
$(U_1,\ldots,U_s,V)\tr$. In particular, for $j=1,\ldots,m$,
there is conditional independence between the response $Y_j$ (observable or not)
and the indicator for response $R_j$, given the latent variables.  
This condition is similar to that of non-ignorable random-coefficient-based drop out 
\citep{little:95}. 

The above assumptions imply that the proposed SEM reduces to a multidimensional IRT model
similar to the model proposed by \cite{bart:07}
for the augmented set of responses $Y_1,\ldots,Y_m,R_1,\ldots,R_m$;
see also \cite{dav:08}. In particular, this model is characterized by the following features:
\begin{itemize}
\item certain response variable $Y_j$ may be missing at random (given also
$R_1,\ldots,R_m$), whereas the variables $R_j$ are always observed;
\item every $Y_j$ depends on a specific latent variable $U_{d_j}$ through a Rasch or a 2PL
parametrization, whereas $R_j$ depends on both 
$U_{d_j}$ and another latent variable $V$, being $d_j$ the dimension measured by item $j$ (i.e., $z_{dj}=1$ if $d=d_j$ and $z_{dj}=0$ otherwise);
\item the latent variables $U_1,\ldots,U_s$ and $V$ have a discrete distribution depending on the individual covariates $X_1,\ldots,X_c$ through a multinomial logistic model. 
\end{itemize}

Consequently, for a vector of observed responses $\b y_{obs}$, which is a subvector
of $(y_1,\ldots,y_m)\tr$, and a realization $\b r=(r_1,\ldots,r_m)\tr$ of $R_1,\ldots,R_m$,
we have the {\em manifest distribution}:
\begin{eqnarray}\label{eq:manifest_cov1}
p({\b y}_{obs},\b r|\b x) &=&  \sum_{h_1=1}^{k_1}\sum_{h_2=1}^{k_2} 
\la_{h_1}(\b x)\pi_{h_2}(\b x)p_{h_1h_2}({\b y}_{obs},\b r|\b x),\\
p_{h_1h_2}({\b y}_{obs},\b r) &=&\prod_{j=1\,(r_j=1)}^mp_{h_1j}^{y_j}
(1-p_{h_1j})^{1-y_j}\prod_{j=1}^m q_{h_2j}^{r_j}(1-q_{h_2j})^{1-r_j},\nonumber
\end{eqnarray}
where the first product is extended to all items $j$ for which a response is observed,
and then $r_j=1$. 
In fact, this model has a number of free parameters equal to
\[
\# \textrm{par} = (k_1+k_2-2)(c+1) +sk_1+k_2 + 2m-(s+1) + I\{a=1\}[3m-(s+1)],
\]
where $s+1$ is the total number of latent variables, 
$a=0$ for the Rasch model, and $a=1$ for to the 2PL model.
In fact, there are $k_1+k_2-2$ regression coefficients for the latent classes for each of the 
$c$ covariates plus one intercept, $sk_1+k_2$ 
ability parameters, $3m-(s+1)$ discriminating parameters (when $a=1$), and $2m-(s+1)$
difficulty parameters.
In fact, $s+1$ difficulty parameters must be constrained to 0 and $s+1$ discriminating parameters must be constrained to 1 to ensure model identifiability.
\section{Likelihood inference}\label {sec:inf}
The parameters of the proposed model, 
which are collected in the vector $\b\eta$, may be estimated through the discrete marginal maximum likelihood approach, making use of the EM algorithm \citep{demp:laird:rub:77}.
For a sample of $n$ subjects, the model log-likelihood is defined as
\[
\ell(\b\eta) = \sum_{i=1}^n \log p({\b y}_{i,obs},\b r_i|\b x_i),
\]
where ${\b y}_{i,obs}$ is the vector of observed responses for subject $i$, $\b r_i$ is the corresponding vector of indicators of response, $\b x_i$ is the vector of covariates for this subject, and $p({\b y}_{i,obs},\b r_i|\b x_i)$ is 
defined in (\ref{eq:manifest_cov1}).

In order to maximize $\ell(\b\eta)$, the EM algorithm
alternates two steps until convergence:
\begin{itemize}
\item[{\bf E-step}:] compute the expected value of the so-called 
{\em complete data log-likelihood} given the current parameter vector; this function
is defined as
\[
\ell^*(\b\eta) = \sum_{h_1=1}^{k_1}\sum_{h_2=1}^{k_2}\sum_{i=1}^n w_{h_1h_2i}
\log[\la_{h_1}(\b x_i)\pi_{h_2}(\b x_i)]+
\sum_{h_1=1}^{k_1}\sum_{h_2=1}^{k_2}\sum_{i=1}^n w_{h_1h_2i} \log p_{h_1h_2}(\b y_{i,obs},\b r_i),
\]
where $w_{h_1h_2i}$ is an indicator variable equal to 1 if subject $i$ belongs to latent
class $h_1$ (with respect to the latent variables $U_1,\ldots,U_s$) and $h_2$ (with respect
to the latent variable $V$) and it is equal to 0 otherwise. Note that in this
case the missing data are referred to values of the latent variables of interest.
Then, the E-step consists of computing the conditional posterior probabilities
\[
\hat{w}_{h_1h_2i} = \frac{\la_{h_1}(\b x_i)\pi_{h_2}(\b x_i)p_{h_1h_2}(\b y_{i,obs},\b r_i)}
{\sum_{t_1=1}^{k_1}\sum_{t_2=1}^{k_2} \la_{t_1}(\b x_i)\pi_{t_2}(\b x_i)p_{t_1t_2}(\b y_{i,obs},\b r_i)},
\]
for $h_1=1,\ldots,k_1$, $h_2=1,\ldots,k_2$, and $i=1,\ldots,n$.
\item[{\bf M-step}:] maximize the expected value of $\ell^*(\b\eta)$ obtained at the E-step with respect to $\b\eta$ so that these parameters are updated. 
This requires to use suitable iterative algorithms of Newton-Raphson type also for the parameters affecting the class weights; see \cite{bart:forc:06} and \cite{bart:07} for details. 
\end{itemize}

The algorithm used for the application is implemented in an {\tt R} function, related to the package {\tt MultiLCIRT}\footnote{Downloadable from \url{http://CRAN.R-project.org/package=MultiLCIRT}.};
this function and the syntax we use are downloadable from  \url{http://www.stat.unipg.it/bartolucci/irt_missing.zip}. 
Note that similar analyses may be performed by means of {\tt Latent GOLD} \citep{latentgold} 
or {\tt Mplus} \citep{mplus} softwares, which allow for multidimensional IRT models, discrete latent traits, MAR data, and individual covariates.

In order to measure the uncertainty about these estimates through standard errors (and confidence intervals), we also implemented a non-parametric bootstrap method \citep{dav:hin:97}, based on drawing with replacement $B$ samples of dimension $n$ from the observed data.
A multidimensional LC IRT model with the same characteristics as the selected model is then estimated for each bootstrap sample and the corresponding quantities of interest are stored. 
In this way, an empirical distribution is obtained, on the basis of which the standard errors  (and  
confidence intervals) may be calculated for every parameter estimate of interest.
Note that the described approach requires some caution. 
In particular, as the ordering of latent classes may change from  
sample to sample, it has to be fixed on the basis of some criteria, 
such as on the basis of the estimated supported points. We also 
outline that the bootstrap resampling procedure is distribution-free;
therefore, it allows us 
to preserve the advantages of our semi-parametric approach and it represents a simple method to estimate standard errors having a complicated distribution.

Finally, it is important to recall  
that the specific LC IRT model that is adopted depends on the chosen  number of latent classes ($k_1$ and $k_2$), the chosen number of latent dimensions ($s$), and the 
chosen parametrization of the conditional response probabilities (Rasch or 2PL).
Therefore, a model selection procedure is usually required in applications involving a certain dataset, unless theoretical considerations do not suggest specific choices, such as a given number of latent classes.
For this aim, we suggest to use a likelihood ratio test when the compared models are nested and the usual regularity conditions are met or, more in general, an information criterion, especially in the presence of non-nested models. 

Different information criteria have been proposed in the statistical literature. Among the best  
known, there are the Akaike Information Criterion \citep[AIC, ][]{aka:73}  and the Bayesian Information Criterion \citep[BIC,][]{sch:78}.  
In particular, we suggest to use BIC, which is based on the minimization of the index:
$BIC = -2\hat{\ell}+\log(n)\#{\rm par}$, where $\hat{\ell}$ is the maximum log-likelihood estimate of the model of interest.

In contrast with AIC, BIC takes into account 
the sample size, so that it uses a larger penalty for the model complexity (measured in terms of number of parameters), and it is asymptotically 
consistent for finite mixture models under certain regularity conditions \citep{keribin:00}.
In particular, BIC tends to select more parsimonious models than  
AIC. For a detailed description of  
AIC and BIC 
we refer the reader to \cite{bur:and:02} and \cite{yang:05}, among others.
In the context of finite mixture models and  
LC models, several studies are 
aimed at comparing the performance of different 
information criteria for model selection.  
We consider, in particular, \cite{fraley:raft:02}, who used BIC for clustering in mixture models showing its satisfactory behavior, and \cite{dias:06}, who compared different selection criteria specifically in the context of LC models; see also \cite{mcla:peel:00}, Ch. 8. \vspace*{-0.25cm}
\section{Simulation study}\label{sec:simula}\vspace*{-0.25cm}
This section illustrates the Monte Carlo simulation experiment we implemented in order to study the finite-sample properties of the maximum likelihood estimates of the proposed model. 
In this way, we can also evaluate empirically the adequacy of the proposed approach in obtaining reliable estimates of the model parameters and in showing the presence of a non-ignorable missing mechanism for the item responses.\vspace*{-0.25cm}
\subsection{Simulation setting}\vspace*{-0.25cm}
The simulation study is based on a model with two latent abilities, $U_1$ and $U_2$, which drive the observed responses, and another latent construct, $V$, which corresponds to the tendency to respond. 
The distribution of these latent variables is based on $k_1 = k_2 = 3$ latent classes
with probabilities affected by two individual covariates, $X_1$ and $X_2$, generated from independent standard normal distributions.
For the sample size we considered $n = 1000, 2000$ and for the number of items $m = 20, 40$. 
The items are equally distributed between $U_1$ and $U_2$, so that the first $m/2$ items are assumed to measure the first latent variable and the second $m/2$ items are assumed to measure the second latent variable.
Moreover, we assumed three types of scenario: (\emph{i}) complete cases (with no missing responses), (\emph{ii}) presence of missing responses depending on $V$, and (\emph{iii}) presence of missing responses depending on $V$ and on $U_1$ and $U_2$. 
In this way, we considered a total of 12 different scenarios, which are listed in Table \ref{tab:scenarios}, under each of which 1000 samples have been simulated.\vspace{5mm} 

TABLE \ref{tab:scenarios}   HERE \vspace{5mm}

All the proposed scenarios assume fixed values of the population parameters.
In particular, the regression coefficients $\phi_{jh}$ of the multinomial logistic model for $U_1$ and $U_2$ are such that covariate $X_1$ has not effect on the distribution of the latent classes ($\phi_{11}=\phi_{12}=0$), whereas covariate $X_2$ has effect only on the logit for class 2 versus class 1 ($\phi_{21} = 1$ and $\phi_{22} = 0$).
Then, the intercepts of the model are chosen so that the average weight of the second latent class ($\bar{\la}_2$) is equal to 0.5, whereas the weights of the other two classes ($\bar{\la}_1$, $\bar{\la}_3$) are equal to 0.25. Finally, regarding the multinomial logistic model for $V$ we assumed $\psi_{jh_2}=\phi_{jh_1}$ for $h_1,h_2=1,2$ and $j=0,1,2$; consequently the average weight of the second class ($\bar{\la}_2$) is again equal to 0.5, whereas the average weights of the other two classes ($\bar{\pi}_1$, $\bar{\pi}_3$) are equal to 0.25.
Moreover, the support points $u_{dh_1}$ and $v_{h_2}$, with $d=1,2$ and $h_1,h_2 =1,2,3$, are chosen so that all latent variables have mean 0 and variance 1 and $U_1$ and $U_2$ are positively correlated, similarly to the situation illustrated in our application based on the Students' Entry Test. 

Concerning the item parameters, the difficulties $\be_j$ correspond to sequence of equally spaced points between $-2$ and 2; the first $m/2$ parameters refer to items of $U_1$ and the remaining $m/2$ to those of $U_2$.
In this way, there are items with increasing difficulties in both dimensions and the discriminating parameters $\al_j$ increase from $1$ to $2$ for $U_1$ and decrease from 2 to 1 for $U_2$.
Therefore, four different types of items are considered: items with high $\be_j$ and high $\al_j$, items with low $\be_j$ and low $\al_j$,  items with high $\be_j$ and low $\al_j$, and items with low $\be_j$ and high $\al_j$.

Finally, we set constant values for $\de_j$ and  $\ga_{2j}$ for $j=1,\ldots,m$, whereas $\ga_{1j}=0$ for $j=1,\ldots,m$ if the missingness depends only on $V$ and $\ga_{1j}=1$ for $j=1,\ldots,m$ if it depends also on $U_1$ and $U_2$.
\subsection{Simulation results}
A bidimensional model with $k_1 = k_2 = 3$ latent classes and a 2PL parametrization were estimated for every simulated sample generated under each of the 12 scenarios. 
Tables \ref{tab:results_1}, \ref{tab:results_2}, and \ref{tab:results_3} show the main 
results of the simulation study, with reference to the estimates of the support points, the regression coefficients, and the item parameters, respectively. 
In particular, we report  the values of the bias and of the root-mean-square-error (RMSE) for each population parameter; for sake of parsimony, we show the average (absolute) value of bias and RMSE for 
the item parameters. 
\vspace{5mm}

TABLES \ref{tab:results_1}, \ref{tab:results_2} AND \ref{tab:results_3}   HERE \vspace{5mm}

We observe generally satisfactorily values of both bias and RMSE. We also observe that bias is approximately constant, regardless of the type of scenario.
On the other hand,  some differences emerge relatively to RMSE. 
In fact, as may be expected, the RMSE decreases when the number of items $m$  
and the sample size $n$ increase, as it is evident by comparing scenarios  1 vs 7, 2 vs 8, and so on for $m$, and 1 vs 4, 2 vs 5, and so on for $n$.  

Concerning the trend of RMSE depending on the changes in the missingness structure,  we observe that the presence or absence of missing item responses does not modify the values of RMSE in case of  regression coefficients, whereas RMSE computed for the item parameters and, though to a lower extent, for the support points is constantly smaller in absence of missingness rather than in its presence (Tables \ref{tab:results_1} and \ref{tab:results_3}, respectively). 
Moreover, some differences emerge by changing the type of missingness dependence. When missing item responses depend only by $V$, RMSE values for the support points of latent distributions (compare scenarios 2 vs 3, 5 vs 6, and 8 vs 9 in Table \ref{tab:results_1}) are higher, those for the regression coefficients tend to be similar, and RMSE values for item parameters are similar or smaller (compare scenarios 5 vs 6 and 11 vs 12 in Table \ref{tab:results_3}) rather than when missing item responses depend also on $U_1$ and $U_2$. 

Finally, a special attention must be paid to the results about the discriminant parameters $\ga_{1j}$.  As shown in Table \ref{tab:results_3}, parameters $\ga_{1j}$ are well estimated both when  the missingness process  depends only  on the latent trait $V$ (scenarios 2, 5, 8, and 11) and when it depends also on the abilities $U_d$, $d=1,2$ (scenarios 3, 6, 9, and 12).  Therefore, the proposed model allows us to obtain evidence about the effect of the latent abilities  on the missing responses and to   properly conclude about the ignorability of the missingness process. In practice, discriminant parameter estimates   $\hat{\ga}_{1j}$ not significantly different from zero for all items denote that the presence of missing responses does not depend on the abilities, whereas values of $\hat{\ga}_{1j}$ significantly different from 0 (for at least some items) denote  that the  missing data provides 
information about the abilities and, accordingly, it cannot  
be ignored. 
\section{Application to Students' Entry Test}\label{sec:applic}\vspace*{-0.25cm}
In the following, we first provide a description of the dataset on which the application is based and then we illustrate the main aspects of the estimated model, with special attention to the results concerning the model selection procedure, the estimated latent structure, and the estimated item parameters, even in comparison with the results obtained assuming ignorable missing responses.\vspace*{-0.25cm}
\subsection{Data description}\label{sec:data}\vspace*{-0.25cm}
The dataset contains the responses provided by 1217 students to an Entry Test for the Faculty of Economics of an Italian university in 2011.
The test is organized in three disciplinary sections concerning the measurement of the same number of abilities represented here
by latent variables: Logic (13 items, measuring latent variable $U_1$), Mathematics 
(13 items, measuring latent variable $U_2$), and Verbal comprehension (10 items,
measuring latent variable $U_3$). Following the proposed approach, 
we also considered a further latent variable, $V$, describing the propensity to answer,
which is assumed to be independent of $(U_1,U_2,U_3)\tr$ given the individual covariates. 
The dataset also contains individual information about three covariates: 
gender (0 = male, 1 = female), type of secondary school diploma (0 = classical or scientific  diploma, 
1  = technical or professional  diploma), and secondary school final mark 
(in 100ths). 

All items are of multiple choice type, with one correct answer and four distractors.
The recorded data are about the correctness or incorrectness of the provided
response. However, since a wrong response penalizes the score more than a response
that is not provided, many missing responses exist in the dataset, whose distribution differs both within the  three disciplinary sections and between the sections themselves. In average, 
the percentage of missing responses is equal to 8.9$\%$ for Verbal comprehension,
14.6$\%$ for Logic, and 38.5$\%$ for Mathematics, whereas the percentage of 
right responses is equal to 73.8$\%$, 59.9$\%$, and 41.1$\%$, respectively. \vspace*{-0.25cm}
\subsection{Model selection}\label{sec:modselec}\vspace*{-0.25cm}
To analyze the data described above, we initially considered the proposed model with $k_1=k_2=3$ latent classes and a Rasch and a 2PL parametrization. 
The first model, denoted by $M_1$, has a maximum-likelihood equal to $\hat{\ell}=-33528.28$ and $BIC=67738.55$. 
For the second model, denoted by $M_2$, we have $\hat{\ell}=-32974.23$ and $BIC=67369.29$.
Therefore, the deviance (likelihood ratio statistic) between the two models is equal to $1108.09$ with $104$ degrees of freedom leading to reject the Rasch parametrization in favor of the 2PL parametrization.
This is also confirmed by the lower value of $BIC$ for $M_2$ than for  
$M_1$. 

Then we fitted  a third model ($M_3$) that is a restricted version of model $M_2$ in which the abilities do not affect the probability to respond, and then in equation (\ref{eq:3}) we have $\ga_{1j}=0$, $j=1,\ldots,m$. 
For model $M_3$ we have $\hat{\ell}=-33256.2$ and $BIC=67677.48$. 
The deviance between models $M_2$ and $M_3$, equal to $563.93$ with 36 degrees of freedom, and the corresponding values of $BIC$ lead to the conclusion that $M_3$ cannot be accepted and then there is evidence that the different abilities affect the probability to respond to the test items. 
In this way we respond to the main scientific question behind the present application, that is, whether there is an influence of the abilities measured by the test items on the probability of responding. 

We recall that the above results are based on the assumption that $k_1=k_2=3$.
The choice $k_1=3$ is motivated by the necessity of distinguishing students admitted, admitted with reserve, and not admitted to university courses in Economics. 
Concerning $k_2$, we verified that BIC suggests $k_2>1$,  giving some evidence about the presence of a distinct latent trait ($V$)  other than the abilities $U_1$, $U_2$, and $U_3$. 
In fact, with $k_1=3$ and $k_2=1$, under the same parametrization of model $M_2$ we obtain $\hat{\ell}=-33989.46$ and $BIC = 69080.07$, which is much higher than the value found with $k_2=3$.

We also verified  that the main conclusions of the study are substantially unaffected by the specific number of latent classes that is selected and then we retained $k_1$ and $k_2$ equal to 3, also to simplify the interpretation of the results.

\subsection{Estimated latent structure}

In order to further study the response mechanism of interest in the present application, in Table \ref{tab:support} we report the distribution of the latent variables, with support points which are denoted by $\hat{u}_{dh_1}^*$ and $\hat{v}_{h_2}^*$, which are standardized so that each latent distribution has mean 0 and variance 1.
In Table \ref{tab:corrtheta} we also report the estimated correlation coefficients  
together with the standard errors based on 199 non-parametric bootstrap samples.
We recall that these estimates are obtained under model $M_2$, that is, the model with 2PL parametrization and covariates affecting the latent distribution, which is based on $k_1=k_2=3$ support points.\vspace{5mm}

TABLES \ref{tab:support} AND \ref{tab:corrtheta}   HERE \vspace{5mm}

Concerning the estimated distribution of abilities measured by the test items, 
we observe that the latent classes are increasingly 
ordered for each dimension. 
Moreover, for each dimension, such a distribution is roughly symmetric, with the
support point for the second class close to 0 (corresponding to the mean) and
the support points for the first and the third classes which are almost equally
distant from 0.
The second class has the highest weight (45.2\%), whereas the other two classes have a similar weight.
On the other hand, for the latent variable representing the tendency to respond, the distribution is far from being symmetric, since there is a small group of subjects, those in the third class, with a high tendency to respond.
The other subjects (87\%) are in the first two classes and have a low or moderate tendency to respond.
Note that the distance between the first and the second class is around one half of the distance between the second and the third. 

Finally, in Table \ref{tab:regression} we report the estimated regression coefficients of the multinomial logistic models which relates the distribution of the latent variables above with the covariates; see assumption (\ref{eq:cov}). \vspace{5mm}

TABLE \ref{tab:regression}   HERE \vspace{5mm}

It is worth noting that all covariates have a significant effect on the distribution
of the latent classes for the abilities, with females and students having classical or scientific high
school diploma that tend to be in classes of lower ability with respect to males
and students having technical or professional school diploma, respectively. It is
also observed that the probability of belonging to the higher ability classes
increases with the final mark. 
Another important aspect that may be observed on the basis of the estimates in Table 
\ref{tab:regression} is that no covariate has a clearly significant effect on the distribution
of the latent variable $V$ and then on the tendency to respond. From the scientific
point of view we then conclude that, differently from the abilities,
this tendency is rather unpredictable on the basis of
the covariates, being related to aspects of the student's personality
that are not directly observable. 
\subsection{Estimated item parameters}
Another relevant 
set of estimates to consider is that of 
the item parameters, which enter in the structural equations (\ref{eq:23}) and (\ref{eq:3}) and then in the logistic equations (\ref{eq:par1}) and (\ref{eq:par2})
for the distribution of the response variables given the latent variables.
These estimates are reported in Table \ref{tab:items} together with 
the bootstrap standard errors.\vspace{5mm}

TABLE \ref{tab:items}   HERE \vspace{5mm}

The first important aspect to consider is that almost all discriminant indices
for the abilities ($\al_j$) are significantly different from 0 and positive, with the exception of two items, which are in two distinct sections (Mathematics and Verbal comprehension).
In particular, the first of these items is the number 20,  for that $\hat{\al}_{20}=-0.022$, whereas the second is the number 29, for that $\hat{\al}_{29}=0.089$. 
These results are rather plausible, since the first of these items has a very low proportion of right answers (5.85$\%$) and the second has a very high proportion of right answers (76.0$\%$) and then they do not provide an adequate discrimination between less and more capable students.
Moreover, the first is the most difficult item according to the results in Table \ref{tab:items} ($\hat{\be}_{20}=1.966$).
On the other hand, the item with the highest discriminant index is the number 24, 
which has an intermediate proportion of right responses (49.0$\%$).
It is also worth noting that many other items have high values of the discriminant index, and then they provide adequate measures of the ability.

On the basis of the results in Table \ref{tab:items} we can also deal with one of the main scientific questions behind the present application.
In particular, once we have established that at least for some items the probability of responding depends on the ability (see previous section about the preference of model $M_2$ compared to model $M_3$), we can verify if this is true for all items or not on the basis of the estimates of the parameters $\ga_{1j}$. 
We note that there are 14 items for which this coefficient is significantly different from 0 and in all these cases
the estimate is positive. Therefore, we conclude that a missing response to one of these items may be interpreted as a sign of low ability level and then this missing response is reasonable to be penalized in scoring the questionnaire. Another relevant aspect is that 10 of these 14 items belong to the section of questionnaire about Mathematics, whereas only 3 belong to the section about Logic and only 1 belongs to the last section about Verbal comprehension.
This means that the dependence of the probability to respond on the ability is particularly
evident for the items about Mathematics, whereas much less evident for those about Verbal
comprehension. 

Finally, the estimates of the discriminant indices $\ga_{2j}$ allow us to conclude that $V$ is an adequate representation of the tendency to respond (independently of the ability level).
In fact, these parameters are significantly different from 0 and have a positive
estimate for 32 items on 36. The interpretation of the single values of these estimates
is similar to that of the estimates of discriminant indices $\al_j$ relating the ability
level to the probability of responding correctly.
\subsection{Comparison with the missing at random approach}\label{sec:MAR}
As suggested by the results about the statistical significance of discriminant parameters $\ga_{1j}$ described in the previous section, we conclude that missing responses are informative about the abilities measured by the test.
To further investigate about this issue we have estimated a multidimensional LC IRT model under the assumption of ignorable missingness and we can compare the results with those obtained through our approach which takes the missingness mechanism explicitly into account.
Here we present the main results related with  the multidimensional LC 2PL model with $k_1=3$ latent classes and $s=3$ latent variables ($U_1$, $U_2$, and $U_3$), being omitted the tendency to answer ($V$).   

Obviously, the MAR approach does not allow to get information about the missingness mechanism and to characterize the latent classes in terms of probability of answering. Moreover, the consequences about the inferential conclusions and about the evaluation of individuals are far from trivial. 

First, even if we do not observe any sensible difference under the estimates of the support points (results are here omitted), the MAR assumption causes a different allocation of individuals among the latent classes. Indeed, ordering the latent classes according to the support points, we note that the estimated average weight for class 2 under MAR assumption is higher than the corresponding value under MNAR assumption ($\hat{\bar{\la}}_{2}$ equals 0.548 and $0.452$, respectively), whereas the estimated average weight for class 3 is lower ($\hat{\bar{\la}}_{3}$ equals 0.189 and 0.280, respectively).

Second, we also note some relevant differences about the estimates of difficulty parameters, $\hat{\be}_j$. Comparing results in the last two columns of  Table \ref{tab:items} (i.e., $\tilde{\al}_j$ and $\tilde{\be}_j$) with values $\hat{\al}_j$ and $\hat{\be}_j$, we observe several values quite different under the two assumptions. In particular,  more unstable estimates are obtained under the MAR approach, as shown by values for items $j=20$ and $j=29$, where the presence of a very high number of missing responses does not allow to reliably evaluate the difficulty of the corresponding items.
Moreover, the estimates for $\be_j$ are less accurate under the MAR approach, being the standard errors always higher when the missingness mechanism is ignored.
\subsection{Conclusions}\vspace*{-0.25cm}
From the scientific point of view, the main conclusions
of the application are the following:

\begin{enumerate}
\item There is a set of items (mostly concentrated in the section about Mathematics)
for which the probability to respond significantly depends on the ability and, therefore, it is reasonable to penalize a missing response to these items in scoring the questionnaire.
\item For the other items (sections about Logic and Verbal comprehension) the probability of responding has a weak dependence on the ability that is measured.
\item The tendency to respond, considered as a distinct latent variable with respect to the abilities, has a distribution with a shape different from that of the measured
abilities.
This distribution is asymmetric, with a reduced number of subjects having a very high tendency to respond.
\item The distribution of the tendency to respond has
a weak dependence on individual covariates, as it is related to 
aspects of the personality that are difficult to be 
predicted on the basis of observable covariates; on the other hand, the same
covariates have a significant effect on the distribution of the abilities.
\item The alternative simple approach based on ignoring the missing responses  modifies the main inferential conclusions about (\emph{i}) the allocation of individuals to the latent classes and (\emph{ii}) the stability and accuracy of the estimates of the difficulty item parameters. 
\end{enumerate}

\small{
\bibliography{biblio}
\bibliographystyle{apalike}}

\newpage

\vspace{1cm}

\begin{table}[!ht]
\begin{center}   
\caption{\em Description of the simulated scenarios by sample size ($n$), number of items ($m$), presence of  missing item responses (yes/no), and type of dependence of missingness on latent traits ($U_1, U_2, V$).}
\label{tab:scenarios}      
\begin{tabular}{cccccrccccc}
\hline\hline
Scenario	&	$n$	&	$m$	&	\multicolumn2c{ missing	item responses	}	&&
Scenario	&	$n$	&	$m$	&	\multicolumn2c{ missing	item responses	}\\
\cline{4-5}\cline{10-11}
	         &		&		&	presence	&	dependence on	&&
   	         &		&		&	presence	&	dependence on	\\
\hline
1	&	1000	&	20	&	no	&	-- 	&& 7	&	1000	&	40	&	no	&	-- 	\\
2	&	1000	&	20	&	yes	&	$V$	&& 8	&	1000	&	40	&	yes	&	$V$	\\
3	&	1000	&	20	&	yes	&	$U_1, U_2, V$	&& 9	&	1000	&	40	&	yes	&	$U_1, U_2, V$	\\
4	&	2000	&	20	&	no	&	-- 	&& 10	&	2000	&	40	&	no	&	-- 	\\
5	&	2000	&	20	&	yes	&	$V$	&& 11	&	2000	&	40	&	yes	&	$V$	\\
6	&	2000	&	20	&	yes	&	$U_1, U_2, V$	&& 12	&	2000	&	40	&	yes	&	$U_1, U_2, V$	\\
\hline													
\end{tabular} 
\end{center}
\end{table}

\begin{table}[!ht]
\begin{center}  
\caption{\em Bias and RMSE for the estimates of the support points of the latent distributions.}  
\label{tab:results_1}      
\begin{tabular}{ll|ccc|ccc|ccc}
\hline\hline
	&		&	\multicolumn3{c|}{Latent class ($h_1, h_2$)}					&	\multicolumn3{c|}{Latent class ($h_1, h_2$)}&	\multicolumn3c{Latent class ($h_1, h_2$)}\\
	\cline{3-11}
	&		&	1	&	2	&	3	&	1	&	2	&	3	&	1	&	2	&	3	\\
\hline
	&		&	\multicolumn3{c|}{Scenario 1}		&	\multicolumn3{c|}{Scenario 2}			&	\multicolumn3c{Scenario 3}			\\
\multirow{2}{*}{$u_{1h_1}$}	&	bias	&	-0.003	&	-0.001	&	-0.002	&	-0.002	&	0.001	&	-0.001	&	-0.008	&	-0.001	&	-0.003	\\
	&	RMSE	&	0.061	&	0.035	&	0.037	&	0.065	&	0.038	&	0.044	&	0.062	&	0.036	&	0.037	\\
\multirow{2}{*}{$u_{2h_1}$}	&	bias	&	-0.001	&	-0.002	&	-0.002	&	0.002	&	-0.001	&	0.002	&	-0.001	&	-0.002	&	0.001	\\
	&	RMSE	&	0.037	&	0.035	&	0.059	&	0.042	&	0.038	&	0.063	&	0.037	&	0.036	&	0.062	\\
\multirow{2}{*}{$v_{h_2}$}	&	bias	&	-- 	&	--	&	--	&	0.002	&	-0.005	&	0.005	&	0.002	&	-0.004	&	0.005	\\
	&	RMSE	&	 --	&	--	&	--	&	0.053	&	0.055	&	0.094	&	0.055	&	0.057	&	0.076	\\
\hline
	&		&	\multicolumn3{c|}{Scenario 4}		&	\multicolumn3{c|}{Scenario 5}			&	\multicolumn3c{Scenario 6}			\\
\multirow{2}{*}{$u_{1h_1}$}	&	bias	&	-0.003	&	0.000	&	-0.001	&	-0.001	&	0.000	&	0.001	&	-0.001	&	0.000	&	0.001	\\
	&	RMSE	&	0.043	&	0.024	&	0.026	&	0.044	&	0.026	&	0.030	&	0.042	&	0.025	&	0.026	\\
\multirow{2}{*}{$u_{2h_1}$}	&	bias	&	0.000	&	-0.001	&	0.001	&	0.000	&	0.001	&	0.003	&	0.001	&	-0.001	&	0.002	\\
	&	RMSE	&	0.026	&	0.024	&	0.042	&	0.032	&	0.026	&	0.043	&	0.027	&	0.024	&	0.043	\\
\multirow{2}{*}{$v_{h_2}$}	&	bias	&	-- 	&	--	&	--	&	0.001	&	-0.003	&	0.003	&	0.001	&	-0.002	&	0.001	\\
	&	RMSE	&	-- 	&	--	&	--	&	0.037	&	0.039	&	0.064	&	0.041	&	0.040	&	0.054	\\
\hline
	&		&	\multicolumn3{c|}{Scenario 7}		&	\multicolumn3{c|}{Scenario 8}			&	\multicolumn3c{Scenario 9}			\\
\multirow{2}{*}{$u_{1h_1}$}	&	bias	&	-0.001	&	0.001	&	0.000	&	0.000	&	0.001	&	0.002	&	-0.002	&	0.001	&	-0.001	\\
	&	RMSE	&	0.056	&	0.031	&	0.028	&	0.058	&	0.032	&	0.032	&	0.056	&	0.029	&	0.028	\\
\multirow{2}{*}{$u_{2h_1}$}	&	bias	&	0.000	&	0.000	&	0.001	&	0.002	&	0.001	&	0.006	&	0.001	&	0.000	&	0.003	\\
	&	RMSE	&	0.029	&	0.030	&	0.057	&	0.032	&	0.032	&	0.061	&	0.028	&	0.029	&	0.054	\\
\multirow{2}{*}{$v_{h_2}$}	&	bias	&	--	&	--	&	--	&	0.002	&	-0.002	&	0.005	&	0.000	&	-0.004	&	0.004	\\
	&	RMSE	&	--	&	--	&	--	&	0.042	&	0.037	&	0.058	&	0.044	&	0.038	&	0.050	\\
\hline
	&		&	\multicolumn3{c|}{Scenario 10}		&	\multicolumn3{c|}{Scenario 11}			&	\multicolumn3c{Scenario 12}			\\
\multirow{2}{*}{$u_{1h_1}$}	&	bias	&	-0.003	&	-0.001	&	-0.001	&	-0.003	&	-0.002	&	0.000	&	-0.004	&	-0.001	&	-0.001	\\
	&	RMSE	&	0.042	&	0.022	&	0.021	&	0.041	&	0.022	&	0.022	&	0.044	&	0.022	&	0.021	\\
\multirow{2}{*}{$u_{2h_1}$}	&	bias	&	-0.001	&	-0.001	&	-0.001	&	-0.002	&	-0.001	&	-0.002	&	-0.001	&	-0.001	&	0.000	\\
	&	RMSE	&	0.020	&	0.022	&	0.040	&	0.022	&	0.022	&	0.041	&	0.020	&	0.023	&	0.043	\\
\multirow{2}{*}{$v_{h_2}$}	&	bias	&	--	&	--	&	--	&	0.001	&	-0.002	&	0.001	&	0.001	&	-0.001	&	0.003	\\
	&	RMSE	&	--	&	--	&	--	&	0.031	&	0.026	&	0.038	&	0.032	&	0.028	&	0.036	\\
\hline
\end{tabular}   
\end{center}
\end{table}

\begin{table}[!ht]
\begin{center}   
\caption{\em Bias and RMSE for the estimates of regression parameters of the constant term, $X_1$, and $X_2$.}  
\label{tab:results_2}    
\footnotesize
\begin{tabular}{ll|cc|cccc|cccc}
\hline\hline
	&		&	\multicolumn2{c|}{Regression parameters}		&	\multicolumn4{c|}{Regression parameters}	&		\multicolumn4{c}{Regression parameters}	\\
		\cline{3-12}
	&		&	$\phi_{j1}$	&	$\phi_{j2}$		&	$\phi_{j1}$	&	$\phi_{j2}$	&	$\psi_{j1}$	&	$\psi_{j2}$	&	$\phi_{j1}$	&	$\phi_{j2}$	&	$\psi_{j1}$	&	$\psi_{j2}$	\\
\hline
	&		&	\multicolumn2{c|}{Scenario 1}	&	\multicolumn4{c|}{Scenario 2}		&	\multicolumn4{c}{Scenario 3}	\\
Const. ($j=0$)	&	Bias 	&	0.003	&	0.008		&	0.004	&	0.003	&	-0.001	&	-0.004	&	0.012	&	0.007	&	-0.001	&	-0.005	\\
	&	RMSE	&	0.093	&	0.103		&	0.098	&	0.105	&	0.121	&	0.177	&	0.094	&	0.105	&	0.127	&	0.170	\\
$X_1$ ($j=1$)	&	Bias 	&	0.000	&	-0.002		&	-0.004	&	-0.004	&	0.001	&	-0.001	&	-0.001	&	-0.003	&	0.000	&	0.003	\\
	&	RMSE	&	0.088	&	0.090	&	0.087	&	0.096	&	0.106	&	0.105	&	0.088	&	0.090	&	0.117	&	0.109	\\
$X_2$ ($j=2$)	&	Bias 	&	0.008	&	0.008	 	&	0.011	&	0.008	&	0.020	&	-0.004	&	0.009	&	0.001	&	0.024	&	-0.006	\\
	&	RMSE	&	0.102	&	0.104		&	0.105	&	0.105	&	0.137	&	0.133	&	0.102	&	0.102	&	0.151	&	0.132	\\
\hline
	&		&	\multicolumn2{c|}{Scenario 4}	&	\multicolumn4{c|}{Scenario 5}		&	\multicolumn4{c}{Scenario 6}	\\
Const. ($j=0$)	&	Bias 	&	0.003	&	0.003		&	0.003	&	-0.001	&	-0.001	&	0.000	&	0.004	&	0.000	&	-0.002	&	-0.002	\\
	&	RMSE	&	0.063	&	0.070	&	0.064	&	0.074	&	0.083	&	0.124	&	0.063	&	0.075	&	0.095	&	0.121	\\
$X_1$ ($j=1$)	&	Bias 	&	-0.001	&	0.001		&	-0.003	&	0.000	&	0.002	&	0.001	&	-0.002	&	-0.004	&	0.001	&	0.000	\\
	&	RMSE	&	0.061	&	0.063	&	0.065	&	0.066	&	0.076	&	0.071	&	0.062	&	0.064	&	0.076	&	0.076	\\
$X_2$ ($j=2$)	&	Bias 	&	0.003	&	0.002		&	0.000	&	0.001	&	0.007	&	-0.001	&	0.004	&	0.002	&	0.011	&	-0.004	\\
	&	RMSE	&	0.071	&	0.070		&	0.074	&	0.075	&	0.093	&	0.094	&	0.072	&	0.071	&	0.101	&	0.095	\\
\hline
	&		&	\multicolumn2{c|}{Scenario 7}	&	\multicolumn4{c|}{Scenario 8}		&	\multicolumn4{c}{Scenario 9}	\\
Const. ($j=0$)	&	Bias 	&	0.000	&	0.001		&	0.003	&	-0.004	&	0.001	&	-0.003	&	0.004	&	-0.002	&	0.008	&	0.003	\\
	&	RMSE	&	0.084	&	0.097		&	0.088	&	0.102	&	0.090	&	0.118	&	0.084	&	0.093	&	0.093	&	0.113	\\
$X_1$ ($j=1$)	&	Bias 	&	0.005	&	0.000	 	&	-0.002	&	0.003	&	-0.002	&	0.000	&	0.002	&	0.000	&	-0.002	&	0.008	\\
	&	RMSE	&	0.084	&	0.089		&	0.086	&	0.090	&	0.090	&	0.098	&	0.087	&	0.094	&	0.092	&	0.098	\\
$X_2$	($j=2$) &	Bias 	&	0.009	&	0.005	 	&	0.011	&	0.004	&	0.009	&	-0.001	&	0.006	&	0.000	&	0.015	&	0.001	\\
	&	RMSE	&	0.094	&	0.103		&	0.098	&	0.099	&	0.105	&	0.107	&	0.094	&	0.098	&	0.114	&	0.113	\\
\hline
	&		&	\multicolumn2{c|}{Scenario 10}	&	\multicolumn4{c|}{Scenario 11}		&	\multicolumn4{c}{Scenario 12}	\\
Const. ($j=0$)	&	Bias 	&	0.003	&	0.003		&	0.003	&	0.004	&	0.001	&	0.002	&	0.005	&	0.005	&	0.001	&	-0.003	\\
	&	RMSE	&	0.061	&	0.070		&	0.060	&	0.069	&	0.068	&	0.081	&	0.063	&	0.074	&	0.067	&	0.083	\\
$X_1$ ($j=1$)	&	Bias 	&	0.001	&	-0.002		&	-0.002	&	0.000	&	0.001	&	0.000	&	0.000	&	0.000	&	-0.001	&	0.000	\\
	&	RMSE	&	0.058	&	0.064	 	&	0.059	&	0.063	&	0.064	&	0.067	&	0.059	&	0.064	&	0.065	&	0.068	\\
$X_2$	($j=2$)&	Bias 	&	0.004	&	0.000	 	&	0.006	&	-0.001	&	0.009	&	0.003	&	0.002	&	0.001	&	0.005	&	-0.002	\\
	&	RMSE	&	0.068	&	0.070	 	&	0.069	&	0.068	&	0.074	&	0.073	&	0.069	&	0.070	&	0.081	&	0.076	\\
\hline
\end{tabular}    
\end{center}
\end{table}

\begin{table}[!ht]
\begin{center} 
\caption{\em Average absolute values of bias and RMSE for the estimates of the item parameters.}  
\label{tab:results_3}     
\begin{tabular}{l|cc|ccccc|ccccc}
\hline\hline
	&	\multicolumn2{c|}{Item parameters} & \multicolumn5{c|}{Item parameters}	&		\multicolumn5{c}{Item parameters}	\\
	\cline{2-13}
	&	$\al_j$	&	$\be_j$	&	$\al_j$	&	$\be_j$	&	$\ga_{1j}$	&	$\ga_{2j}$	&	$\de_j$	& $\al_j$	&	$\be_j$	&	$\ga_{1j}$	&	$\ga_{2j}$	&	$\de_j$	\\
\hline
	&	\multicolumn2{c|}{Scenario 1}			&	\multicolumn5{c|}{Scenario 2} & \multicolumn5{c}{Scenario 3}		\\
bias 	&	0.009	&	0.004	&	0.015	&	0.007	&	0.002	&	0.009	&	0.008	&	0.013	&	0.006	&	0.006	&	0.007	&	0.013	\\
RMSE	&	0.110	&	0.111	&	0.134	&	0.127	&	0.087	&	0.109	&	0.100	&	0.138	&	0.128	&	0.108	&	0.114	&	0.109	\\
\hline
	&	\multicolumn2{c|}{Scenario 4}			&	\multicolumn5{c|}{Scenario 5} & \multicolumn5{c}{Scenario 6}		\\
bias 	&	0.005	&	0.003	&	0.006	&	0.005	&	0.002	&	0.005	&	0.005	&	0.007	&	0.004	&	0.004	&	0.005	&	0.005	\\
RMSE	&	0.076	&	0.077	&	0.090	&	0.087	&	0.062	&	0.076	&	0.071	&	0.096	&	0.090	&	0.077	&	0.080	&	0.077	\\
\hline
	&	\multicolumn2{c|}{Scenario 7}			&	\multicolumn5{c|}{Scenario 8} & \multicolumn5{c}{Scenario 9}		\\
bias 	&	0.009	&	0.005	&	0.011	&	0.006	&	0.002	&	0.007	&	0.006	&	0.013	&	0.006	&	0.006	&	0.007	&	0.008	\\
RMSE	&	0.105	&	0.107	&	0.122	&	0.122	&	0.082	&	0.099	&	0.097	&	0.131	&	0.123	&	0.104	&	0.103	&	0.107	\\
\hline
	&	\multicolumn2{c|}{Scenario 10}			&	\multicolumn5{c|}{Scenario 11} & \multicolumn5{c}{Scenario 12}		\\
bias 	&	0.004	&	0.003	&	0.005	&	0.003	&	0.001	&	0.003	&	0.004	&	0.005	&	0.004	&	0.003	&	0.003	&	0.005	\\
RMSE	&	0.074	&	0.076	&	0.085	&	0.085	&	0.058	&	0.069	&	0.068	&	0.091	&	0.088	&	0.073	&	0.072	&	0.076	\\
\hline
\end{tabular}     
\end{center}
\end{table}

\begin{table}[!ht]
\begin{center}    
\caption{\em Estimated 
support points for the ability distribution ($\hat{u}^*_{h_1d}$) and for the tendency
to respond ($\hat{v}^*_{h_2}$) and corresponding average weights ($\hat{\bar{\la}}_{h_1}$, $\hat{\bar{\pi}}_{h_2}$).
The support points are standardized so as to have mean 0 and variance 1.}
\label{tab:support}    
\begin{tabular}{lrrr}
\hline\hline
	&		\multicolumn3{c}{Latent class $h_1, h_2$} 	\\
\cline{2-4}
 	             &	\multicolumn1c{1}	&	\multicolumn1c{2}	&	\multicolumn1c{3}\\
\hline
Ability 1 ($\hat{u}^*_{1h_1}$)	&	-1.471 &  0.116 &  1.217\\
Ability 2 ($\hat{u}^*_{2h_1}$)	&	-1.189 & -0.205 &  1.466\\
Ability 3 ($\hat{u}^*_{3h_1}$)	&	 -1.603 &  0.356  & 0.956\\
Probability ($\hat{\bar{\la}}_{h_1}$)	&	0.267 & 0.452 & 0.280\\
\hline													
Tendency ($\hat{v}_{h_2}^*$)	&	-1.046 & 0.051 & 2.296\\
Probability ($\hat{\bar{\pi}}_{h_2}$)	&	0.305  &  0.569 &  0.126\\
\hline													
\end{tabular}   
\end{center}
\end{table}

\begin{table}[!ht]
\begin{center}    
\caption{\em Estimates of correlation coefficients between latent traits, with non-parametric bootstrap standard errors in parentheses.}
\label{tab:corrtheta}   
\begin{tabular}{l|cccc}
\hline\hline
	&	$U_1$	&	$U_2$	&	$U_3$	\\
\hline
$U_1$	& 1.000 & 0.957 &  0.975\\
        & (0.000) & (0.075) & (0.038)\\
$U_2$   & 0.957 & 1.000 &  0.869\\
        & (0.075) & (0.000) & (0.133)\\
$U_3$   & 0.975 &  0.869 &  1.000	\\
        & (0.038) & (0.133) & (0.000)\\
\hline
\end{tabular}    
\end{center}
\end{table}

\begin{table}[!ht]
\begin{center}   
\caption{\em Estimates of regression coefficients ($\hat{\phi}_{jh_1}$, $\hat{\psi}_{jh_2}$, $h_1, h_2 = 1,2$), with non-parametric bootstrap standard errors in parentheses. Estimates of the parameters which are different from 0 at $5\%$
significance level are indicated by a star ($*$).}
\label{tab:regression}       
\begin{tabular}{l|rrrrrrrr}
\hline\hline
	&	\multicolumn1c{$\hat{\phi}_{j1}$}	&	\multicolumn1c{$\hat{\phi}_{j2}$}	&	
	\multicolumn1c{$\hat{\psi}_{j1}$}	&	\multicolumn1c{$\hat{\psi}_{j2}$}	\\
\hline											
Constant	($j=0$)  &		-3.107$^*$ &  -6.171$^*$  &   -0.246\:\:  &  -0.922\:\: \\
 & (1.165)  & (1.302)  & (0.672)  & (0.955)\\
Gender ($j=1$)	&		-1.331$^*$ &   -2.243$^*$  &   -0.276\:\:  &  -0.092\:\:\\
 & (0.281)  & (0.429)  & (0.229)  & (0.364)\\
Diploma ($j=2$)	&		-1.168$^*$ &   -3.866$^*$  &    0.465\:\:  &   0.163\:\:\\
                & (0.555)  & (0.687)  & (0.239)  & (0.396)\\
Final mark ($j=3$)	&    0.068$^*$ &     0.116$^*$  &     0.010\:\:  &   0.000\:\:\\
 & (0.016)  & (0.019)  & (0.010)  & (0.022)\\\hline			
\end{tabular} 
\end{center}
\end{table}

\begin{table}[!ht]
\begin{center}    
\caption{\em Estimates of the item parameters under MNAR assumption ($\hat{\al}_j$, $\hat{\be}_j$, $\hat{\ga}_{1j}$, $\hat{\ga}_{2j}$, $\hat{\de}_j$) and under  MAR assumption ($\tilde{\al}_j$, $\tilde{\be}_j$). 
The latent traits are standardized and non-parametric bootstrap standard errors are in parentheses; estimates of the parameters different from 0 at the $5\%$ significance level are indicated by a star ($*$).}
\label{tab:items}       
{\footnotesize
\begin{tabular}{r|rrrrrrrrrrrr}
\hline\hline
$j$	& \multicolumn1c{$\hat{\al}_j$}	& \multicolumn1c{$\hat{\be}_j$}	
& \multicolumn1c{$\hat{\ga}_{1j}$} & \multicolumn1c{$\hat{\ga}_{2j}$}	
& \multicolumn1c{$\hat{\de}_j$}	& \multicolumn1c{$\tilde{\al}_j$}	& \multicolumn1c{$\tilde{\be}_j$}\\
\hline
  1 & 0.466$^*$ (0.129) & -2.382$^*$ (0.110) & 0.291\:\: (0.276) & 1.611$^*$ (0.485) & -3.964$^*$ (0.348) & 0.575$^*$(0.176)& -2.413$^*$(0.132)\\ 
2	&	 0.639$^*$ (0.138)	&	 -2.168$^*$ (0.129)	&	 0.362\:\: (0.271)	&	 1.267$^*$ (0.424)	&	-3.722$^*$ (0.289)	&	1.409$^*$(0.193)	&	-0.513\:\:(0.900)	 \\
3	&	 0.981$^*$ (0.210)	&	 -2.881$^*$ (0.197)	&	 0.550$^*$ (0.188)	&	 1.376$^*$ (0.253)	&	-3.159$^*$ (0.173)	&	1.954$^*$(0.392)	&	-2.431\:\:(2.512)	 \\
4	&	 0.508$^*$ (0.100)	&	 -0.185$^*$ (0.073)	&	 0.070\:\: (0.238)	&	 1.644$^*$ (0.361)	&	-2.414$^*$ (0.191)	&	0.716$^*$(0.105)	&	-0.762\:\:(1.056)	 \\
5	&	 0.836$^*$ (0.130)	&	 -0.840$^*$ (0.085)	&	 0.379\:\: (0.200)	&	 1.703$^*$ (0.379)	&	-2.221$^*$ (0.191)	&	2.012$^*$(0.184)	&	0.008\:\:(1.660)	 \\
6	&	 0.855$^*$ (0.144)	&	 -2.103$^*$ (0.126)	&	 0.820$^*$ (0.265)	&	 1.811$^*$ (0.478)	&	-3.501$^*$ (0.338)	&	1.890$^*$(0.292)	&	-2.052\:\:(2.327)	 \\
7	&	 0.314$^*$ (0.109)	&	 0.826$^*$ (0.091)	&	 0.022\:\: (0.149)	&	 1.132$^*$ (0.226)	&	-0.334$^*$ (0.090)	&	0.389$^*$(0.118)	&	2.032\:\:(1.730)	 \\
8	&	 0.781$^*$ (0.124)	&	 -2.034$^*$ (0.115)	&	 -0.045\:\: (0.340)	&	 2.119\:\: (4.697)	&	-4.508\; (4.554)	&	1.881$^*$(0.192)	&	-0.531\:\:(1.467)	 \\
9	&	 1.106$^*$ (0.149)	&	 -0.894$^*$ (0.095)	&	 0.354\:\: (0.245)	&	 1.681$^*$ (0.424)	&	-3.432$^*$ (0.275)	&	2.096$^*$(0.258)	&	-1.848\:\:(2.380)	 \\
10	&	 0.732$^*$ (0.120)	&	 -1.411$^*$ (0.088)	&	 1.295$^*$ (0.249)	&	 1.676$^*$ (0.294)	&	-1.773$^*$ (0.178)	&	0.737$^*$(0.175)	&	-1.852$^*$(0.843)	 \\
11	&	 0.201$^*$ (0.084)	&	 -0.314$^*$ (0.061)	&	 0.139\:\: (0.200)	&	 1.460$^*$ (0.311)	&	-1.637$^*$ (0.147)	&	0.610$^*$(0.101)	&	1.300$^*$(0.472)	 \\
12	&	 0.772$^*$ (0.092)	&	 0.219$^*$ (0.068)	&	 -0.168\:\: (0.219)	&	 1.300$^*$ (0.287)	&	-2.391$^*$ (0.159)	&	1.293$^*$(0.114)	&	0.151\:\:(1.398)	 \\
13	&	 1.106$^*$ (0.136)	&	 0.615$^*$ (0.089)	&	 -0.123\:\: (0.196)	&	 1.488$^*$ (0.312)	&	-2.021$^*$ (0.162)	&	1.382$^*$(0.168)	&	-3.160\:\:(2.129)	\\\hline
14	&	 1.049$^*$ (0.156)	&	 -0.904$^*$ (0.130)	&	 0.426$^*$ (0.192)	&	 1.269$^*$ (0.200)	&	0.325$^*$ (0.098)	&	0.999$^*$(0.145)	&	-0.954$^*$(0.116)	 \\
15	&	 1.107$^*$ (0.144)	&	 0.153\:\: (0.150)	&	 0.683$^*$ (0.182)	&	 0.855$^*$ (0.110)	&	0.799$^*$ (0.075)	&	0.877$^*$(0.170)	&	-0.730$^*$(0.223)	 \\
16	&	 0.714$^*$ (0.109)	&	 -0.664$^*$ (0.087)	&	 0.650$^*$ (0.151)	&	 0.600$^*$ (0.101)	&	-0.852$^*$ (0.069)	&	0.379$^*$(0.105)	&	-1.723$^*$(0.159)	 \\
17	&	 0.401$^*$ (0.103)	&	 0.304$^*$ (0.084)	&	 0.745$^*$ (0.153)	&	 0.618$^*$ (0.099)	&	-0.243$^*$ (0.070)	&	0.349$^*$(0.105)	&	1.289$^*$(0.410)	 \\
18	&	 0.438$^*$ (0.100)	&	 -0.124\:\: (0.073)	&	 0.240\:\: (0.169)	&	 0.966$^*$ (0.143)	&	-0.533$^*$ (0.083)	&	0.402$^*$(0.098)	&	-0.087\:\:(0.199)	 \\
19	&	 0.835$^*$ (0.176)	&	 -1.885$^*$ (0.133)	&	 0.620$^*$ (0.110)	&	 0.603$^*$ (0.102)	&	-0.555$^*$ (0.077)	&	0.472$^*$(0.155)	&	-3.398$^*$(0.375)	 \\
20	&	 -0.022\:\: (0.237)	&	 1.966$^*$ (0.188)	&	 -0.491\:\: (0.173)	&	 1.032$^*$ (0.252)	&	0.856$^*$ (0.102)	&	0.017\:\:(0.086)	&	118.802(187.769)	 \\
21	&	 0.316$^*$ (0.087)	&	 -0.279$^*$ (0.079)	&	 0.847$^*$ (0.192)	&	 1.128$^*$ (0.190)	&	-1.079$^*$ (0.106)	&	0.210$^*$(0.090)	&	-0.591\:\:(0.877)	 \\
22	&	 0.914$^*$ (0.289)	&	 -2.935$^*$ (0.246)	&	 0.197\:\: (0.132)	&	 0.652$^*$ (0.156)	&	-1.843$^*$ (0.098)	&	0.546$^*$(0.195)	&	-4.492$^*$(0.524)	 \\
23	&	 1.122$^*$ (0.284)	&	 -2.273$^*$ (0.193)	&	 0.517$^*$ (0.206)	&	 1.329$^*$ (0.308)	&	-2.725$^*$ (0.188)	&	1.074$^*$(0.173)	&	-2.194$^*$(0.209)	 \\
24	&	 1.303$^*$ (0.194)	&	 -1.049$^*$ (0.128)	&	 0.910$^*$ (0.192)	&	 1.175$^*$ (0.229)	&	-1.094$^*$ (0.107)	&	0.994$^*$(0.163)	&	-1.722$^*$(0.122)	 \\
25	&	 0.698$^*$ (0.106)	&	 -0.147\:\: (0.076)	&	 0.325$^*$ (0.144)	&	 0.949$^*$ (0.175)	&	-0.924$^*$ (0.095)	&	0.499$^*$(0.109)	&	-1.338$^*$(0.148)	 \\
26	&	 1.034$^*$ (0.129)	&	 -0.592$^*$ (0.096)	&	 0.816$^*$ (0.178)	&	 0.488$^*$ (0.106)	&	-1.031$^*$ (0.084)	&	0.966$^*$(0.118)	&	-0.626$^*$(0.113)	\\\hline
27	&	 0.719$^*$ (0.106)	&	 -1.414$^*$ (0.090)	&	 0.331\:\: (0.259)	&	 1.864\:\: (5.539)	&	-3.873\; (5.172)	&	0.789$^*$(0.101)	&	-1.447$^*$(0.092)	 \\
28	&	 0.482$^*$ (0.126)	&	 -2.599$^*$ (0.124)	&	 0.256\:\: (2.604)	&	 0.906\:\: (7.230)	&	-5.106\; (8.057)	&	0.345$^*$(0.132)	&	-4.326\:\:(2.338)	 \\
29	&	 0.089\:\: (0.100)	&	 -2.021$^*$ (0.091)	&	 -0.243\:\: (0.222)	&	 1.452$^*$ (0.351)	&	-2.360$^*$ (0.175)	&	0.209$^*$(0.101)	&	-8.465\:\:(64.210)	 \\
30	&	 0.660$^*$ (0.165)	&	 -3.171$^*$ (0.168)	&	 0.500\:\: (0.294)	&	 1.476$^*$ (0.512)	&	-4.276$^*$ (0.455)	&	0.754$^*$(0.162)	&	-3.314$^*$(0.500)	 \\
31	&	 0.658$^*$ (0.101)	&	 -1.961$^*$ (0.111)	&	 0.079\:\: (7.143)	&	 1.020 (14.208)	&	-5.850 (19.052)	&	0.468$^*$(0.101)	&	-2.911$^*$(0.276)	 \\
32	&	 0.702$^*$ (0.166)	&	 -3.028$^*$ (0.170)	&	 0.858$^*$ (0.278)	&	 1.965$^*$ (0.487)	&	-3.959$^*$ (0.378)	&	0.982$^*$(0.143)	&	-2.625$^*$(0.398)	 \\
33	&	 0.404$^*$ (0.109)	&	 -2.211$^*$ (0.094)	&	 0.257\:\: (0.311)	&	 1.996$^*$ (0.533)	&	-3.809$^*$ (0.393)	&	0.432$^*$(0.118)	&	-3.401$^*$(1.106)	 \\
34	&	 0.432$^*$ (0.097)	&	 -0.527$^*$ (0.068)	&	 -0.007\:\: (0.292)	&	 1.931$^*$ (0.389)	&	-2.116$^*$ (0.197)	&	0.324$^*$(0.095)	&	-0.852$^*$(0.148)	 \\
35	&	 0.127\:\: (0.074)	&	 -0.579$^*$ (0.069)	&	 -0.055\:\: (0.221)	&	 1.696$^*$ (0.404)	&	-2.710$^*$ (0.235)	&	0.169$^*$(0.068)	&	-2.147\:\:(10.596)	 \\
36	&	 0.287$^*$ (0.094)	&	 -0.003\:\: (0.069)	&	 -0.050\:\: (0.302)	&	 2.362$^*$ (0.467)	&	-2.213$^*$ (0.226)	&	0.294$^*$(0.092)	&	0.901$^*$(0.363)	\\\hline
\end{tabular}}
\end{center}
\end{table}

\end{document}